# THE EFFECT OF DISCONTINUOUS INJECTION ON PARTICLE BACKFLOW IN PNEUMATIC CONVEYING SYSTEMS

Otome Obukohwo[1], Andrew Sowinski[1], Poupak Mehrani[1], Holger Grosshans[2,3]

[1] Department of Chemical and Biological Engineering, University of Ottawa, Ottawa, Canada.

[2] Analysis and Simulation in Explosion Protection, Physikalisch-Technische Bundesanstalt, Braunschweig, Germany.

[3] Institute of Apparatus and Environmental Technology, Otto von Guericke University of Magdeburg, Magdeburg, Germany

**Correspondence:**

Andrew Sowinski

Department of Chemical and Biological Engineering, University of Ottawa, Ottawa, Canada.

Email: andrew.sowinski@uottawa.ca

# ABSTRACT

Pneumatic conveying is used in many process industries to transport dry, granular, and powdered solids. The triboelectrification of particles during conveying causes particle agglomeration, spark discharges, and disruptions in particle flow making particles move upstream against the fluid flow. The effect of frequency of particle injection on particle backflow is studied using CFD-DEM simulations. Conveying flow in a square-shaped duct with fluid frictional Reynold's number equal to 180, particle Stokes number equal to 8, and individual particle charge equal to 504 fC, is simulated with different injection frequencies. The proportion of particles moving upstream is found to increase as the delay period between injections increases, and the effect of the length of the injection period is minimal. Further, particles moving upstream are situated in low-drag zones at the corners of the duct where the electrostatic force dominates. In conclusion, the delay period between discontinuous injections plays a major role in particle backflow. The findings of the article are important for industrial processes with discontinuous injection of particles with a risk of particle accumulation within the conveying boundary.

# NOMENCLATURE

| Variable | Description | Unit |
|---|---|---|
| $\mathbf{u}$ | Fluid velocity vector | m s$^{-1}$ |
| $\rho_f$ | Fluid density | kg m$^{-3}$ |
| $p$ | Dynamic pressure | kg m$^{-1}$ s$^{-2}$ |
| $\nu$ | Kinematic viscosity | m$^2$ s$^{-1}$ |
| $F_s$ | Momentum source term | m s$^{-2}$ |
| $\rho_p$ | Particle density | kg m$^{-3}$ |
| $\omega$ | Local particle volume | m$^3$ |
| $N$ | Number of particles within a local volume | |
| $\mathbf{f}_{fl}$ | Aerodynamic acceleration | m s$^{-2}$ |
| $\mathbf{f}_g$ | Gravitational acceleration | m s$^{-2}$ |
| $\mathbf{f}_{coll}$ | Collisional acceleration | m s$^{-2}$ |
| $\mathbf{f}_{el}$ | Electrostatic acceleration | m s$^{-2}$ |
| $C_d$ | Drag coefficient | |
| $\mathbf{u}_{rel}$ | Particle velocity relative to fluid velocity | m s$^{-1}$ |
| $Re_p$ | Particle Reynold's number | |
| $r$ | Particle radius | m |
| $\mathbf{g}$ | Gravitational constant | m s$^{-2}$ |
| $Q$ | Particle charge | C |
| $\mathbf{E}$ | Electric field strength | kg m C$^{-1}$ s$^{-2}$ |
| $\varphi_{el}$ | Electric potential | kg m$^2$ C$^{-1}$ s$^{-2}$ |
| $\rho_{el}$ | Electric charge density | C m$^{-3}$ |
| $\varepsilon$ | Electric permittivity of solid-gas mixture | F m$^{-1}$ |
| $t_1$ | Injection period | s |
| $t_2$ | Delay period | s |
| $t_{ref}$ | Reference time for injection decision | s |

# 1 INTRODUCTION

Pneumatic conveying is the transport of dry granular or powdered solids using a fluid, usually gas. Air is used as the conveying gas unless special conditions/requirements are to be considered (reactivity requirements, explosive or fire hazards).[1]

Pneumatically conveyed particles become electrostatically charged due to collisions among particles and, collisions between particles and the conveying tube walls. This charging process is called triboelectrification, tribocharging or contact electrification. [2] Pneumatic conveying tribocharges particles the most, compared to other powder handling processes (see FIGURE 1). [3] The effects of particle triboelectrification include particle agglomeration on conveying walls, electric discharges in silos, and dust explosions in conveying systems. [4]

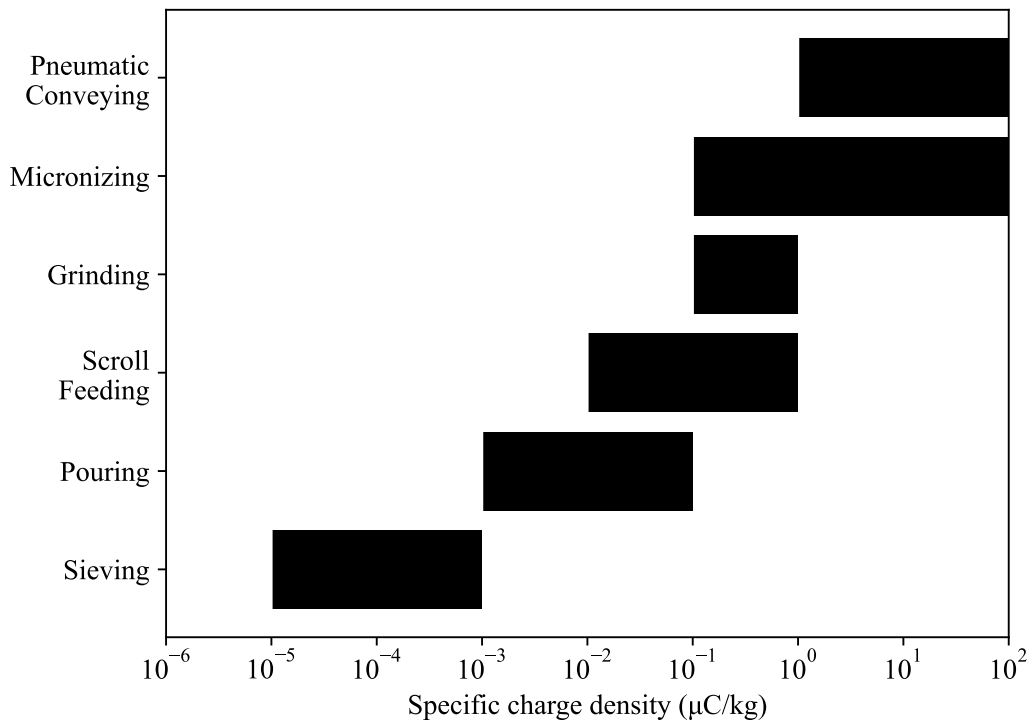

**FIGURE 1** Typical magnitudes of charge generated on particles in industrial processes. [3]

Particle triboelectrification is essential for some industrial processes: electrophotography, powder separation, tomography, and printing [5–7] but, in processes focused on particle flow, triboelectrification leads to loss of material and increased maintenance costs.[8–12]

Nimvari *et al.*[13] where 54 μm dehydrated silica particles were conveyed in nitrogen, in a horizontal stainless-steel conveying tube with a glass section at the outlet of the tube. Backflow of particles during discontinuous injection and the electrostatic charge on the particles was suggested to be responsible for the backflow. Another case is the work of Lim *et al.*[14] where smaller (2.8 μm) polypropylene particles were simulated in air through vertical and 45° inclined pipes with conveying flow against gravity using Computational Fluid Dynamics-Discrete Element Method (CFD-DEM). Particle backflow occurred when the electrostatic field strength in the pipe crossed a certain minimum threshold, again suggesting that electrostatic charge was responsible. A final case is our previous work [15] where we simulated continuous and discontinuous particle injections of 15 μm particles into a horizontal square duct with the aim of qualifying the role of electrostatic charging in particle backflow and determining the conditions necessary for backflow. We found that discontinuous injection of particles and an electrostatic Stokes number ($St_{el}$) greater than $8\times10^{-3}$ were necessary for particle backflow. For clarity: the electrostatic Stokes number is a non-dimensional measure of the ratio of the characteristic timescale of the electrostatic forces to the particle's inertia. In our previous work, an $St_{el}$ value equal to $8\times10^{-3}$ corresponded to a charge of 5.04 fC per particle. Discontinuous injection allows particles to experience upstream acceleration due to the repulsive electrostatic force from other like-charged particles. Discontinuous (batch) injection is used in multiple industries; one being the polyethylene (PE) industry where catalyst particles are injected in batches into the reactor[16] to control the residence time of the particles in the reactor. Particle backflow in such a system could affect the reactor efficiency less particles than intended get into the reactor. This example highlights a need to qualify the effect of the characteristics of discontinuous injection on particle backflow.

This paper reports the results of Direct Numerical Simulation (DNS) - DEM simulations aimed at qualifying the effect of frequency of injection on particle backflow. The mechanism of particle backflow is examined in detail and the frequency of injection, defined as the duration of injection (injection period) and the duration between injections (delay period), is varied, and the effect of this variation is studied.

Numerical simulations are preferred for this paper due to the precision and ease of measurement they offer; for example, they offer precise control and measurement of conveying conditions (particle concentration, particle size, particle charge, etc.) and frequency of injection. Currently, it is impossible to accurately assign or measure a charge magnitude on a single particle among a cluster of moving particles. There are experimentations of single particle charging and neutralization [17,18], but these methods are effective only for completely neutralizing or saturating a particle. Further, in the context of charge measurement - the Faraday cage is the most widely used device, but it only provides net charge measurements and does not properly capture the distribution of charge magnitudes. There are sensors that provide charge distributions with a limitation on particle number density.[19] Finally, numerical simulations allow for better visualization and analysis of data from any desired timepoint in the entire conveying process, which is yet another important requirement for this study.

## 2 METHODOLOGY

This section summarizes the computational methods used and the setup of the simulations. Particle Flow Simulation in Explosion Protection (pafiX) was used to simulate the conveying systems reported in this paper. A summarized explanation for the solver procedures in pafiX is provided in section 2.1. A detailed explanation for pafiX has been published by the authors of the software: Grosshans et al..[20] pafiX has also been extensively validated across multiple published reports. [5,20,21]

### 2.1 Computational Solver

The solver consists of three parts coupled together: the Navier-Stokes equation for fluid flow, Newton's second law of motion for particle movement, and Gauss's law for the electrostatic field.

2.1.1 Fluid motion defined by the Navier-Stokes equation

The fluid phase of the simulations is a fully resolved DNS simulation. The motion of the fluid, according to the Navier-Stokes equation for incompressible flows, is defined as follows:

$$\nabla \cdot \boldsymbol{u} = 0 \tag{1}$$

$$\frac{\partial \boldsymbol{u}}{\partial t} + (\boldsymbol{u} \cdot \nabla)\boldsymbol{u} = -\frac{1}{\rho_\text{f}}\nabla p + \nu\nabla^2\boldsymbol{u} + F_\text{s} \tag{2}$$

where $\boldsymbol{u}$ is the fluid velocity vector, $\rho_\text{f}$ is the fluid density, $p$ is the fluid dynamic pressure, and $\nu$ is the fluid kinematic viscosity.

The source term, $F_s$, represents the momentum transfer from the particles to the fluid and is defined as follows:

$$F_s = -\frac{\rho_p}{\rho_f}\omega \sum_{i=1}^{N} f_{fl,i} \tag{3}$$

where $\rho_p$ is the particle density, $\omega$ represents the local particle volume within a control volume containing $N$ particles, $f_{fl,i}$ represents the sum of all fluid forces acting on particle $i$.

### 2.1.2 Particle motion defined by Newton's second law

The particle phase is simulated using DEM. The positions of the particles are tracked in a Lagrangian framework and Newton's second law of motion is solved for each particle as follows:

$$\frac{du_p}{dt} = f_{fl} + f_g + f_{coll} + f_{el} \tag{4}$$

where $f_{fl}$ is the acceleration due to aerodynamic forces, $f_g$ is the gravitational acceleration, $f_{coll}$ is the acceleration due to particle-particle or particle-wall collisions, and $f_{el}$ is the acceleration due to electrostatic forces.

The acceleration due to the aerodynamic forces is calculated as follows:

$$f_{fl} = -\frac{3\rho}{8\rho_p r} C_d |u_{rel}| u_{rel} \tag{5}$$

$$u_{rel} = u_p - u \tag{6}$$

where $C_d$ is the particle drag coefficient, and $u_{rel}$ is the particle velocity relative to the fluid velocity.

$C_d$ is calculated as a function of the particle Reynolds number ($Re_p$) by the following:

$$Re_p = \frac{2|u_{rel}|r}{\nu} \tag{7}$$

$$C_d = \begin{cases} \frac{4}{Re_p}\left(6 + Re_p^{\frac{2}{3}}\right), & Re_p \leq 1000 \\ 0.424, & Re_p > 1000 \end{cases} \tag{8}$$

where $r$ represents the particle radius.

The gravitational acceleration is calculated as a function of the gravitational constant (**g**) as follows:

$$\boldsymbol{f}_{\text{g}} = \left(1 - \frac{\rho}{\rho_{\text{p}}}\right)\boldsymbol{g} \qquad (9)$$

Particle-particle collisions are modelled using a variant of the hard-sphere approach called the ray casting method to detect collisions between particles (see **FIGURE 2**). The criteria for anticipating collisions between two particles in order of ascending computational costs are:

1. Check whether both particles are in the same or adjacent computational cell(s) with the assumption that the particle velocity is similar to the fluid velocity and the particles do not traverse more than one computational cell per time-step.
2. Check if the angle between the vector connecting the particles' centroids ($\boldsymbol{z}_{12}$) and the relative velocity vector ($\boldsymbol{v}_1^*$) is acute. An acute angle suggests that particles are on a collision course.
3. Determine whether the particles will collide if they continue with their current velocity.
4. Determine whether the anticipated collision takes place in the following time increment.

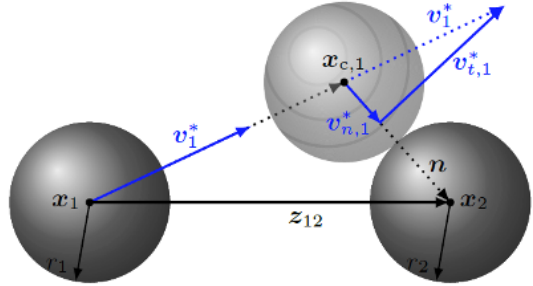

**FIGURE 2** Parameters used in the ray casting collision detection in the rest frame of particle 2. Adapted from Grosshans et al..[20]

If any criterion is not fulfilled, the following criteria area are ignored, and the particles are propagated with $\boldsymbol{f}_{\text{coll}} = 0$. If all criteria are fulfilled, an oblique collision is executed.

The acceleration due to electrostatic forces acting on the particles is calculated as follows:

$$\boldsymbol{f}_{\text{el}} = \frac{Q\boldsymbol{E}}{m_{\text{p}}}. \qquad (10)$$

where $Q$ is the charge on the particle, $m_\text{p}$ is the mass of the particle, and $\boldsymbol{E}$ is the strength of the electric field around the particle.

$\boldsymbol{E}$ is calculated using a combination of Gauss's law and Coulomb's point charge force. The effect of charged particles in the same computational cell is calculated as a Coulombic interaction while the effect of particles outside the computational cell is considered using Gauss's law. This scheme has been thoroughly tested and validated.[22] The electric field, dictated by Coulomb's law, is calculated as follows:

$$E(\boldsymbol{d}) = \frac{Q\boldsymbol{d}}{4\pi\varepsilon|\boldsymbol{d}^3|} \qquad (11)$$

where $\boldsymbol{d}$ is the vector connecting the centers of the interacting particles and $\varepsilon$ is the electric permittivity of the solid-gas mixture ($8.85 \times 10^{-12}$ F/m). Gauss's law, as a function of the electric potential $\varphi_\text{el}$, by the following:

$$\boldsymbol{E} = -\nabla\varphi_\text{el} \qquad (12)$$

So, Gauss's law is reduced to the following:

$$\nabla^2\varphi_\text{el} = -\frac{\rho_\text{el}}{\varepsilon}. \qquad (13)$$

where $\rho_\text{el}$ is the electric charge density. The described method for calculating particle charge effect is a simplified method that does not consider polarization effects or heterogenous charge distribution on particle surfaces. There are other methods that do consider these effects and may offer greater accuracy in simulation results. [23]

## 2.2 Simulation Setup

All simulations in the study were bounded in a square duct (see **FIGURE 3**). For clarity, the physics of fluid-particle interactions in a square duct are distinct from channel, or cylindrical pipe flows, due to cross-sectional vortices. Thus, the results reported in this paper are specifically for square-shaped ducts. The frictional Reynolds number of fluid in all simulations is 180. The particle properties are shown in Table 1. The charge value chosen was found to cause particle backflow in the current flow conditions.[15] For clarity, the charge of each particle is preset, equal, and constant; there is no tribo-charging happening during the simulation. This is intentionally done to ensure that the differences between simulations are due only to the change in the frequency of injection. These simulations, though unrealistic, will provide insight into the effects of particle injection frequency on particle backflow.

Table 1. Properties of particles in simulation.

| Property | Value | Unit |
|---|---|---|
| Diameter | 15 | μm |
| Density | 9480 | kgm$^{-3}$ |
| Charge per particle | 504 | fC |

The fluid kinematic viscosity and density were 1.46 m$^2$s$^{-1}$ and 1.2 kgm$^{-3}$, respectively. The particle and fluid properties were chosen to obtain a Stokes number of 8. The simulated flow conditions are representative of typical pneumatic conveying systems with the fluid friction Reynolds number being at the lower end of the operating range and the particle Stokes number being on the upper end of the typical operating range.

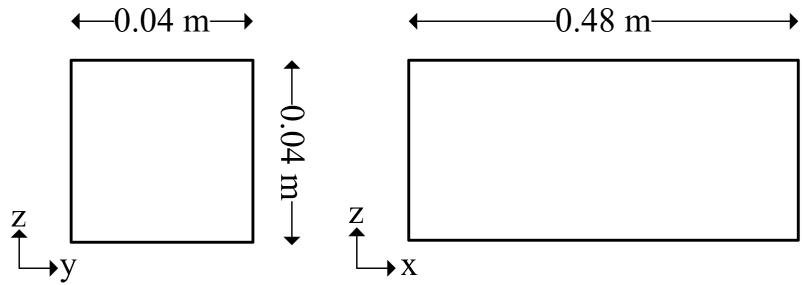

**FIGURE 3** Schematic of simulation boundary.

The computational domain was discretized by 512 × 144 × 144 computational cells, in the x, y, and z directions respectively. A stretched grid was used to provide more definition to the areas closest to the walls. A grid independence study has been performed for the same boundary shape and dimensions [24] and the optimal grid size was found to be 480 × 140 × 140 cells. The simulations in this study had a coarser grid; so, the results are not limited by the grid size.

Gravitational force was neglected in the simulations as the focus of the study was on the interplay of fluid mechanics and electrostatics. The duct walls were assumed to be conductive and grounded, consequently, zero-Dirichlet boundary conditions were applied to the electric potential of the system.

Discontinuous injection of particles, at a rate of 5×10⁵ particles per second, was modelled to resemble square waves with two defining parameters: $t_1$ and $t_2$, defined as the injection period of a pulse and, the delay period between two consecutive pulses, respectively. For clarity, discontinuous injection profiles are not always square waves, but a square wave was chosen for these simulations because it is the ideal form of discontinuous injection.

The mathematical representation of particle injection is shown in Equation (13) and Equation (14).

$$\dot{N}_{\text{in}}(t) = \begin{cases} 5 \times 10^5 \text{ (s}^{-1}), & t - t_{\text{ref}} < t_1 \\ 0, & t - t_{\text{ref}} > t_1 \end{cases} \tag{13}$$

At any time $t$, $t_{\text{ref}}$ is calculated as follows:

$$t_{\text{ref}} = \left\lfloor \frac{t}{t_1 + t_2} \right\rfloor (t_1 + t_2) \tag{14}$$

where the operator $\lfloor x \rfloor$ denotes the floor function that returns the largest integer less than $x$.

Each $t_1$ value (1.38, 2.75, 5.5, 8.25, 11, 16.5) was simulated with every $t_2$ value (2.75, 5.50, 11, 22, 44, 88, 176, ∞). All injection and delay period values are in milliseconds.

## 3 DISCUSSION OF RESULTS

Analysis begins with the comparison of proportion of particles moving upstream shown in **FIGURE 4**. The proportion of particles moving upstream is the ratio of the number of particles with a negative velocity to the total number of particles in the boundary. The proportion of upstream-moving particles increases slowly and steadily with $t_2$ with a significant jump from 176 ms to ∞ for all injection periods besides 16.5 ms. This proportion increases with the delay period, but there is no apparent effect of the injection period. The increase with the delay period is consistent with the results of our previous study. [15] Discontinuous injection is necessary for particle backflow because the backflow is a result of the electrostatic repulsion among like-charged particles in the cluster. As the delay period reduces, the mean space between the clusters reduces and the flow approaches continuous flow. So, any single particle will experience a similar magnitude of repulsion from both upstream and downstream directions, reducing the likelihood of backflow.

The delay period between injections influences the number of dense clusters of particles in the duct. A higher cluster count implies a lower expected spacing between the clusters which reduces the likelihood of particles moving upstream. [15] FIGURE 5 shows the number of clusters present in the duct after 0.19 s for all simulations, the average spacing between clusters, and side views of the duct with different cluster counts.

The cluster spacing increases with an increase in either delay or injection periods or both. It is directly correlated with the length of the delay period, and a longer injection period allows the particle cluster to dissipate more before the introduction of the next cluster into the duct. Nevertheless, the effect of the injection period on the cluster spacing is significantly less than that of the delay period. The cluster count decreases with an increase in either delay or injection periods or both. A longer injection or delay period results in less injections per unit time and consequently less cluster s simultaneously present in the duct at any given time.

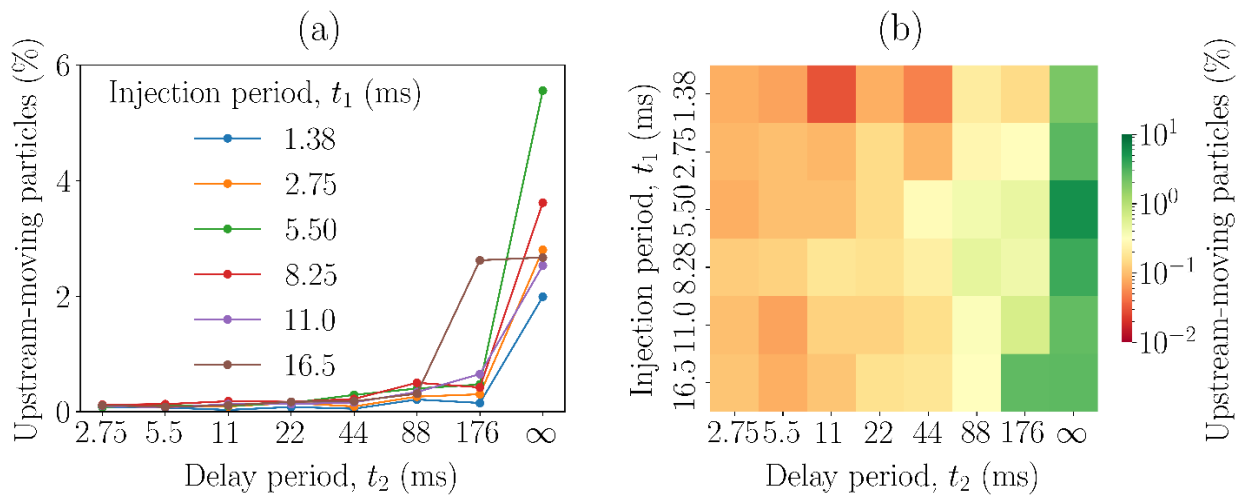

**FIGURE 4** Proportion of particles moving upstream as a function of injection and delay periods shown as a line plot (A) and as a heatmap (B). More particles move upstream as the length of the delay period increases; the length of the injection period has no apparent effect.

The simulations with the lowest cluster count (1 cluster) and highest cluster spacing have the highest proportions of upstream-moving particles. These include all simulations with an infinite delay period and the simulation with injection and delay periods equal to 16.5 ms and 176 ms respectively, as previously observed in **FIGURE 4**. The physical effects of injection and delay periods is shown in **FIGURE 5**C-D. A longer injection period allows for more particles and denser clusters, while the longer delay period allows for larger cluster spacing.

Further analysis of the evolution of particle velocities also highlights the effect of cluster spacing on particle backflow. For example, the temporal evolution of particle velocities for the simulation with injection period equal to 1.38 ms and an infinite delay period is shown in **FIGURE 6**. The cluster is injected at the start of the simulation ($t = 0.05$ s) and the particles follow the velocity profile of the conveying gas (centerline velocity = 3.65 m/s) hence, most particles have a positive velocity. At 0.10 s, the particles in front of the cluster accelerate forward, due to the repulsive electrostatic forces, toward the outlet of the duct. At 0.15 s and 0.20 s, the faster moving particles exit the duct, and the maximum particle velocity decreases. Simultaneously, at 0.10 s, the particles at the tail of the cluster decelerate and the minimum particle velocity decreases through to 0.20 s, as well as the number of slow-moving particles.

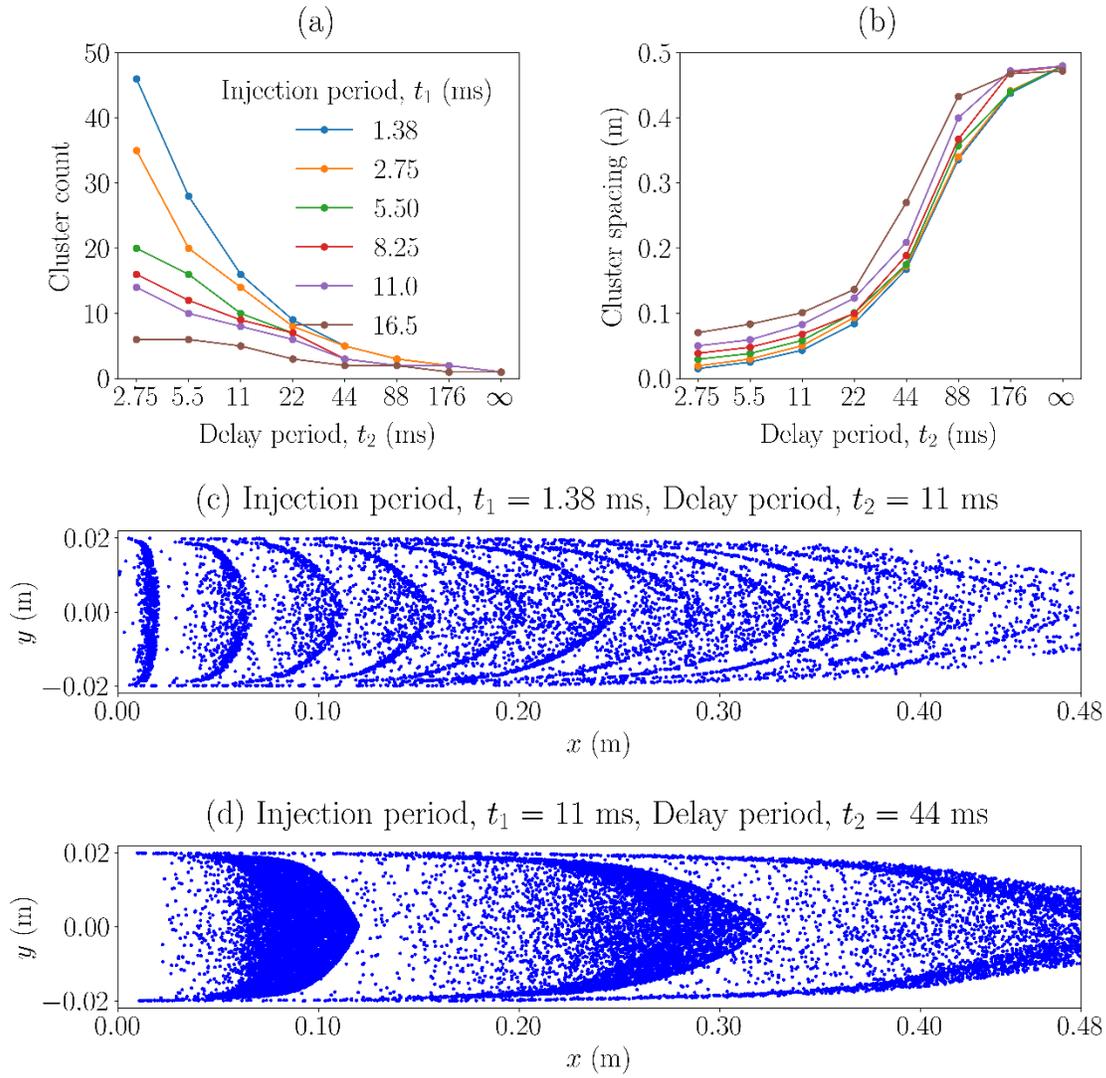

**FIGURE 5** (A) Number of clusters present and (B) the maximum space between clusters as a function of the delay period for all simulations. Side view of tube from after 0.15 seconds of simulation for simulations with injection periods and delay periods equal to (C) 1.38 ms and 11 ms, and (D) 11 ms and 44 ms.

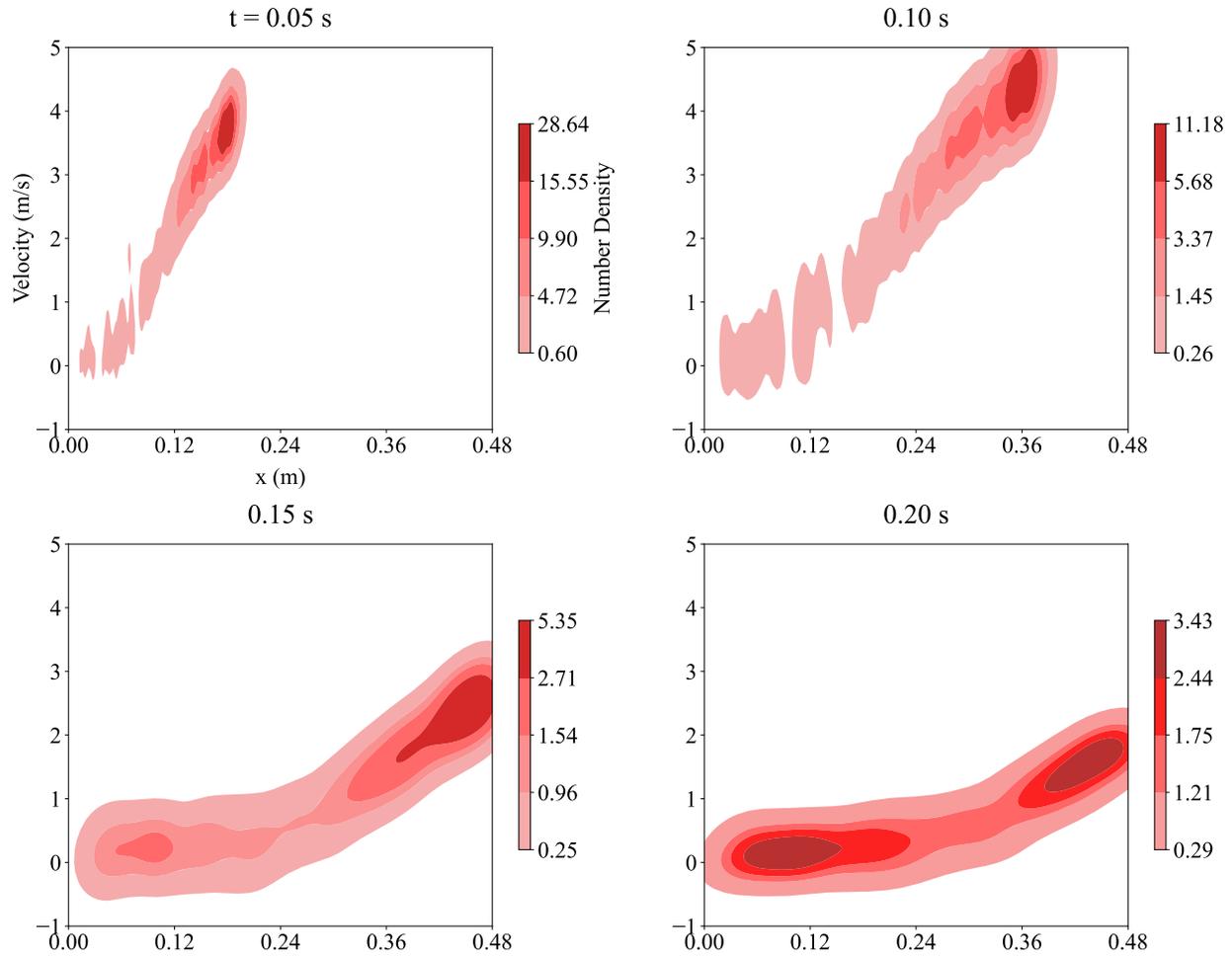

**FIGURE 6** Time-progression of velocity of particles shown, as a probability density plot, as a function of the position along the $x$-axis for simulation with injection period equal to 1.38 ms and an infinite delay period.

A comparison with velocity profiles of other simulations (**FIGURE 7**) buttresses the effect of cluster spacing: the evolution of the velocity profile tends toward the single cluster case as the cluster spacing increases. The velocity density profiles of simulations with shorter delay periods do not develop into the bimodal shape that suggests significant backflow. Instead, newly introduced fast-moving particles represent most of the distribution. In the simulation with a delay period of 176 ms, the bimodal backflow shape is almost fully developed before the new pulse.

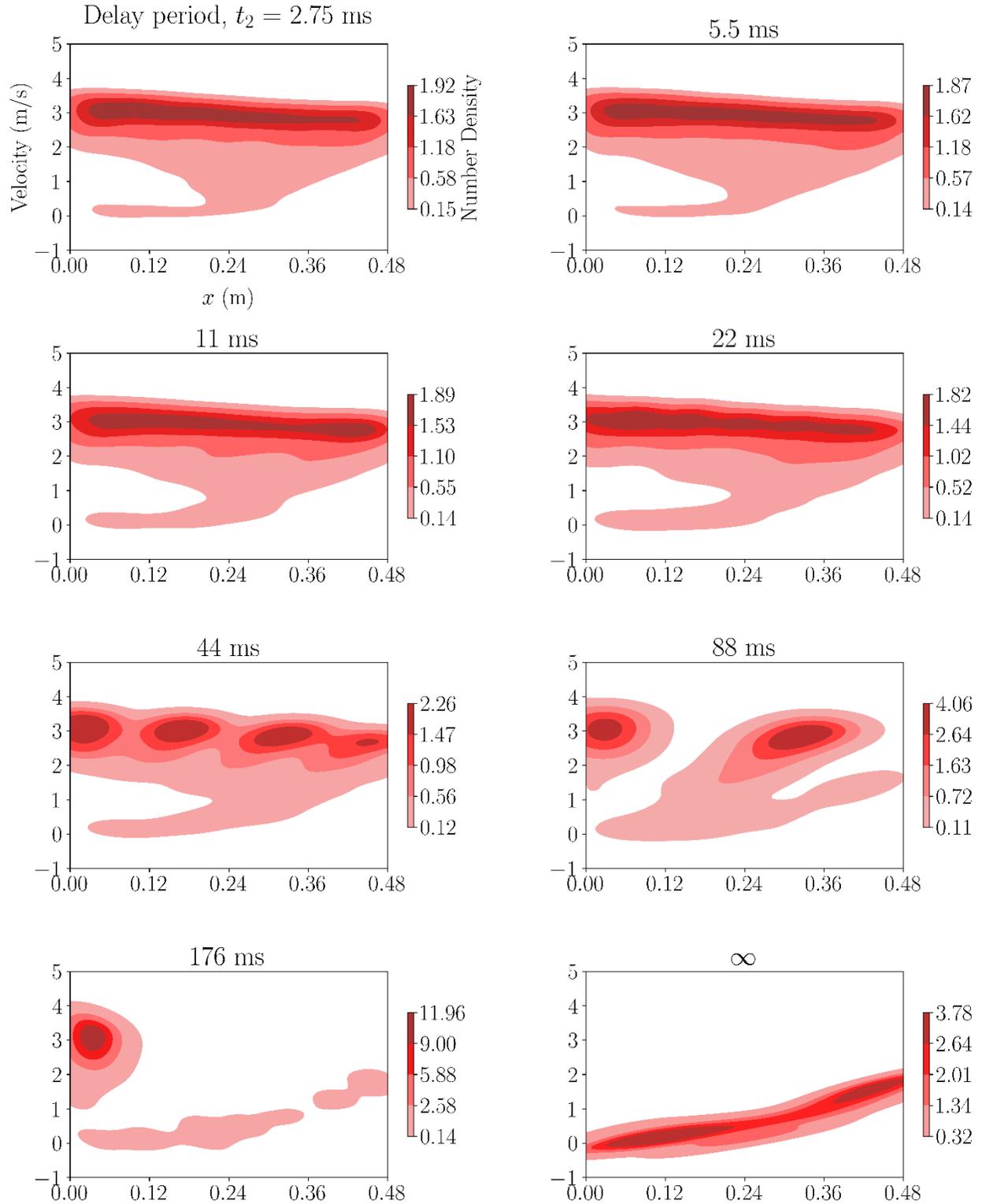

**FIGURE 7** Probability density plots of particle velocity as a function of position along the $x$-axis for all simulations ($t$ = 0.18 s) with an injection period equal to 1.38 ms.

On a final note, it is important to identify the positions of the upstream moving particles. To this end, the streamwise electrostatic and drag accelerations on particles are plotted as a function of the proximity of particles to the walls (**FIGURE 8**) for the simulation with an injection period equal to 5.5 ms and an infinitely long delay period; only one-half of the duct is shown due to the symmetry across the center line. [24]

The magnitudes of the electrostatic acceleration (**FIGURE 8**A) confirms the repulsive push between particles - the scatterplot shows a line of symmetry along $f_{el,x} = 0$ ms$^{-2}$; suggesting that particles experiencing a net positive electrostatic push are accompanied by particles experiencing a similar net negative electrostatic push. However, a comparison of both plots in Figure 8 reveals that the ratio of the electrostatic to drag acceleration is very low: $f_{el,x}$ is 4 orders of magnitude lower than $f_{d,x}$; this ratio has also been reported by Grosshans. [5] The force balance on the particles dictates that the conditions for backflow due to electrostatic repulsion are:

(i) the electrostatic acceleration must oppose the fluid flow: $f_{el,x} < 0$ ms$^{-2}$ if the flow is in positive $x$ direction,

(ii) and the magnitude of the electrostatic acceleration must be greater than the magnitude of the drag acceleration: $|f_{el,x}| > |f_{d,x}|$.

So, the upstream moving particles must be in regions close to the duct wall, where the magnitude of the drag force is low enough for electrostatic forces to dominate. **FIGURE 9** confirms the location of upstream moving particles in the areas with the least fluid velocity and consequently, the least drag force dominance: the corners of the duct.

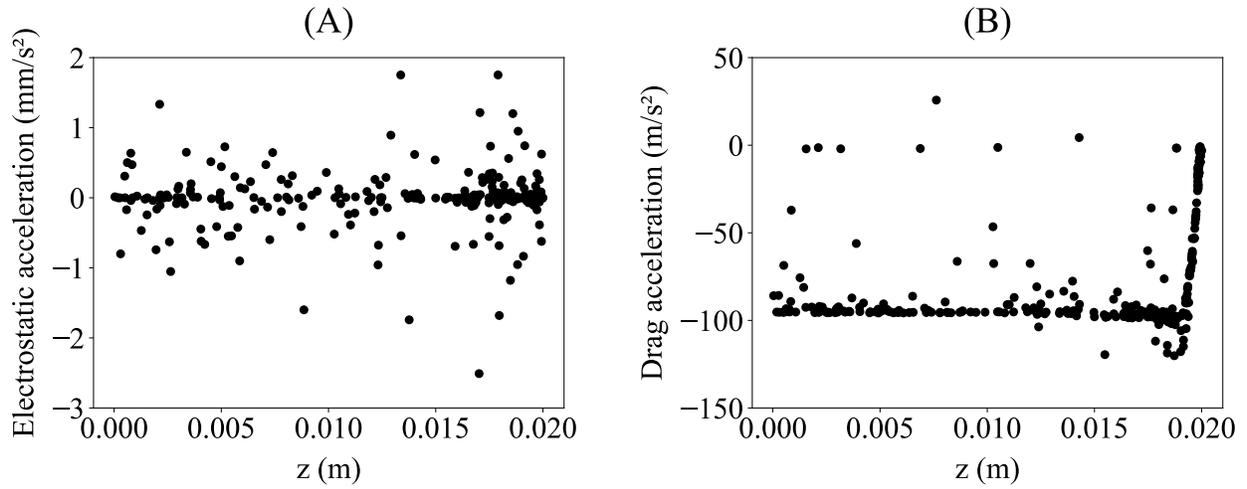

**FIGURE 8** Electrostatic acceleration (A) and drag acceleration (B) as a function of position along the z-axis.

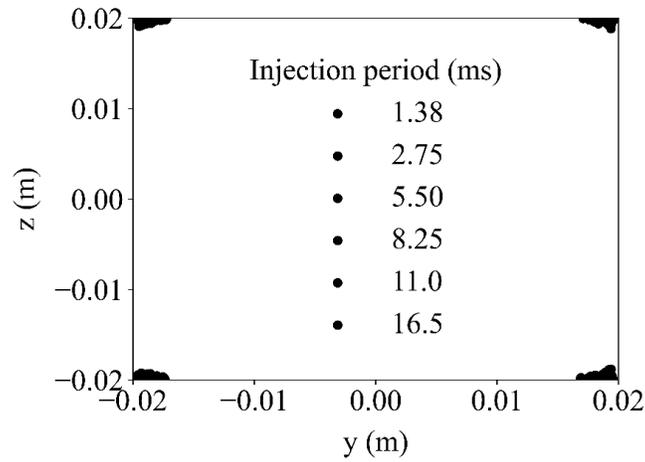

**FIGURE 9** Location of upstream moving particles for all simulations with an infinitely long delay period between injections after 0.19 seconds of conveying (flow into the page).

# CONCLUSIONS

The findings of this paper reveal a new interaction between particle flow properties and tribo-electrification, specifically the effect of discontinuous particle injection on particle backflow in pneumatic conveying systems. The delay period between injections influences the space between particle clusters and consequently, the incidence of particles experiencing decelerative electrostatic forces. The length of the injection period has no significant effect on the likelihood of particles moving upstream. Even with the necessary conditions, upstream-moving particles must overcome drag forces and so are situated in low drag zones at the corners of a conveying duct.

Understanding these effects is relevant for safety in powder flow systems, and for optimizing industrial powder flow processes where discontinuous injections of particles are necessary. The mechanisms of the many effects of triboelectrification are still not completely known; further analysis of particle backflow in systems with polydisperse particle size and charge may yet provide a deeper understanding of the backflow mechanism.

# ACKNOWLEDGEMENT

This work was supported by the National Sciences and Engineering Research Council of Canada (NSERC), the PTB Guest Researcher Fellowship Grant and the European Research Council (ERC) under the European Union's Horizon 2020 research and innovation program (grant agreement No. 947606 PowFEct).